\documentclass{aa}

\begin{document}

\thesaurus{09(09:07:1;09:03:1;08:06:2;02:05:1)}

\title{Neutrino decay and the thermochemical
equilibrium of the interstellar medium}

\author{N\'estor M. S\'anchez D.\inst{1,2}
\and    Neyda Y. A\~{n}ez P.\inst{1}}

\offprints{N. S\'anchez}
\mail{nestor@orion.ciens.luz.ve}

\institute{Departamento de F\'{\i}sica,
           Facultad Experimental de Ciencias,
           Universidad del Zulia,
           Maracaibo, Venezuela
\and       Centro de Astrof\'{\i}sica Te\'{o}rica,
           Facultad de Ciencias,
           Universidad de los Andes,
           M\'{e}rida, Venezuela}

\date{Received 23 September 1999 / Accepted 23 December 1999}

\authorrunning{S\'anchez \and A\~nez}
\titlerunning{Thermochemical equilibrium of the interstellar medium}
\maketitle

\begin{abstract}

We calculate the thermochemical equilibrium of the diffuse
interstellar medium, including ionization by a photon flux
$F_{\nu}$ from neutrino decay. The main heating mechanism
considered is photoelectrons from grains and PAHs. For
the studied
range of $F_{\nu}$ values,
there always exists
two regions of stability (a warm and a cold phase) that can
coexist in equilibrium if
the thermal interstellar pressure is
between a maximum value ($P_{max}$) and a minimum
value ($P_{min}$). High $F_{\nu}$ values
($\sim 10^4-10^5$ cm$^{-2}$ s$^{-1}$)
can be consistent with observed interstellar pressures
only if more efficient
sources are heating the gas.
It is shown that a neutrino flux increase (due, for example,
to an increase in the supernova explosion rate)
may stimulate the condensation of cold gas by decreasing
$P_{max}$ below the interstellar pressure value.

\keywords{ISM: general -- ISM: clouds -- 
stars: formation -- elementary particles}

\end{abstract}

\section{Introduction}

The diffuse interstellar medium (ISM) is observed to be
inhomogeneous with cold ($T\la 10^2$ K)
clouds embedded in a warmer ($T\sim 10^4$ K) intercloud
gas (see, for example, the review of Kulkarni \& Heiles 1988).
The
theoretical explanation for this structure was
provided by Field et al. (1969) who showed
that two thermally stable phases can coexist in pressure
equilibrium over a limited range of pressures, close
to those observed in the ISM. Since the two-phase
model of Field et al. (1969), the thermal and ionization
equilibrium of the ISM and its stability have been
studied by many authors including different heating
and ionizing processes
(Black 1987; Kulkarni \& Heiles 1988).
The response of the ISM to variations in
physical processes and parameters are of interest
both to a better understanding of the ISM 
behavior and because of its central role in star
formation and galaxy evolution models. Effects on the
ISM equilibrium of variations in, for example, X-ray
and far UV radiation fields, cosmic ray ionization and
metal abundance have been studied by several authors
(Shull \& Woods 1985;
Parravano 1987;
Wolfire et al. 1995;
Parravano \& Pech 1997).
In this work, we are
interested in a particular process: the flux of
ionizing photons coming from the radiative decay
of neutrinos (Sciama 1990). Notwithstanding 
this is a speculative theory, Sciama has argued
in a set of papers (Sciama 1993, 1995, 1997a, 1997b, 1998)
that it can explain the widespread ionization far
from the galactic disk and many other observational results.
The simplification in Sciama's work is to assume a
temperature of $\sim 10^4$ K without explicitly
solving the thermal equilibrium. The former was made by
Dettmar \& Schulz (1992) who showed that
heat input associated with neutrino decay is too
small to account for the observed ISM temperature.
However,
their conclusion that neutrino decay cannot be a dominant
source of ionization could be mistaken, as was pointed
out by Sciama (1993),
because they
neglected the existence of other known heating
mechanisms. Our goal here is to analize, in a more
complete and self-consistent way,
the effect
of ionization due to neutrino decay on the thermochemical
equilibrium of the ISM. 

In Sect. 2 we discuss the physical processes included in 
this work and provide the basic equations.
Sect. 3 is dedicated to a discussion of the results, and
the main conclusions are 
summarized in Sect. 4.

\section{Basic equations}

In order to analize the effect of
neutrino decay photons
on the thermochemical equilibrium,
a simple model for the ISM is used. The included
cooling mechanisms are: a) cooling by collisions
of electrons with C$^+$, Si$^+$, Fe$^+$, 
O$^+$, 
S$^+$ and N ($\Lambda_e$); b) cooling by collisions
of neutral hydrogen with 
Si$^+$, Fe$^+$ and C$^+$ ($\Lambda_H$);
and c) cooling due to Ly-$\alpha$
excitation by electrons ($\Lambda_{Ly}$). All the cooling rates and
the relative abundances were 
taken from Dalgarno \& McCray (1972).
We consider the following heating mechanisms:\\
a) Interaction of cosmic ray with hydrogen atoms and electrons:
\begin{eqnarray}
\Gamma_{cr}= \zeta_{cr} \left[ 5\times10^{-12} (1+\Phi) n
(1-\chi)+
5.1\times10^{-10} n_e \right] \nonumber \\
{\rm ergs\ cm^{-3}\ s^{-1}}\ ,
\end{eqnarray}
where $n = n({\rm HI})+n({\rm HII})$
is the total number density of hydrogen,
$n_e$ is the number density of electrons and
$\chi = n({\rm HII})/n$ is the ionization degree of hydrogen.
The number of secondary ionizations
($\Phi$) was taken from Dalgarno
\& McCray (1972), and the primary ionization rate is assumed to be
$\zeta_{cr}=10^{-17}\ {\rm s^{-1}}$ (Spitzer 1978;
Black et al. 1990; Webber 1998).\\
b) Heating by ${\rm H}_2$ formation on dust grains (Spitzer 1978):
\begin{equation}
\Gamma_H=4.4\times10^{-29} n^2
(1-\chi)\ {\rm ergs\ cm^{-3}\ s^{-1}}\ .
\end{equation}
c) Photoelectric heating from small grains and
PAHs (Bakes \& Tielens 1994):
\begin{equation}
\Gamma_{pe}=10^{-24} \epsilon
G_o n\ {\rm ergs\ cm^{-3}\ s^{-1}}\ ,
\end{equation}
where the heating efficiency ($\epsilon$) is given by
\begin{eqnarray}
\epsilon & = & \frac{4.87\times10^{-2}}{\left[ 1+
4\times10^{-3}\left( G_o T^{1/2}/n_e \right) ^{0.73} \right]} +
\nonumber \\
& & \frac{3.65\times10^{-2} \left( T/10^4 \right)^{0.7}}{\left[ 1+
2\times10^{-4}\left( G_o T^{1/2}/n_e \right) \right]}\ ,
\end{eqnarray}
and $G_o$ is the far UV field normalized to its solar
neighborhood value.

We only consider ionization/recombination for hydrogen. The
recombination rate is given by
\begin{equation}
X^+ = \chi n_e \alpha (T)\ {\rm s^{-1}}\ ,
\end{equation}
where $\alpha (T)$ is
the recombination coefficient to all states except the
ground one, and it was taken from Spitzer (1978). The rate of
ionization by cosmic rays is given by
\begin{equation}
X_{cr}^{-} = \zeta_{cr} (1+\phi) (1-\chi)\ {\rm s^{-1}}\ .
\end{equation}
We also use the simple analytic fits provided by
Wolfire et al. (1995) to estimate the ionization
($X_{XR}^{-}$) and heating ($\Gamma_{XR}$) due to
the soft X-ray background as functions of the column
density ($N_w$) and the electron fraction ($n_e/n$).
In this work we adopt $N_w = 10^{19} {\rm cm^{-2}}$.

In addition to the above
sources of ionization,
we also consider the photons
produced by neutrino decay.
The ionization due to this mechanism can
be written in the form (Sciama 1990):
\begin{equation}
X_{\nu}^{-} = F_{\nu} \sigma (1-\chi)\ {\rm s^{-1}}\ ,
\end{equation}
where $\sigma=6.3\times10^{-18}\ {\rm cm^2}$ is
the absorption cross section of hydrogen and $F_{\nu}$
is the flux of hydrogen-ionizing photons produced by neutrino decay.
In this work $F_{\nu}$ is a free parameter, although Sciama (1997a)
estimated that a value of
$F_{\nu} \simeq 3\times10^4$ photons cm$^{-2}$ s$^{-1}$
is necessary to produce an electron density
$n_e \sim 0.05$ cm$^{-3}$ in the intercloud medium. The
most recent (but still uncertain) estimation was
$F_{\nu} \la 10^5$ photons cm$^{-2}$ s$^{-1}$ (Sciama 1998).

\section{Results and discussion}

The thermochemical equilibrium is calculated by solving
simultaneously the equations
$\Lambda = \Gamma$ and $X^{+} = X^{-}$,
where
$\Lambda=\Lambda_e+\Lambda_H+\Lambda_{Ly}$
is the total cooling rate, 
$\Gamma=\Gamma_{cr}+\Gamma_H+\Gamma_{pe}+\Gamma_{XR}$ is
the total heating rate, and
$X^{-}=X^{-}_{cr}+X^{-}_{XR}+X^{-}_{\nu}$ is
the total ionization rate.
Fig.~1a shows the equilibrium pressure-density relations
for $G_o = 1$ and
for three
different values of $F_{\nu}$ ($0$, $10^2$ and $10^4$ 
cm$^{-2}$ s$^{-1}$).
The corresponding electron fractions ($n_e/n$) are showed in Fig.~1b,
where it can be seen that, as expected, $n_e/n$ increases as
$F_\nu$ increases. Most of the ionization for the cases
$F_\nu > 10$ cm$^{-2}$ s$^{-1}$ is
due to photons coming from neutrino decay,
and thus the neutrino
decay is a very efficient ionization mechanism.
An increase in $F_\nu$ (and the consequent increase in
the electron density) enhances the cooling
by electron collisions ($\Lambda_e$).
Additionally, the dominant
heating mechanism is always photoelectrons from
grains and PAHs ($\Gamma_{pe}$), which almost it is
not affected by the flux $F_\nu$. Consequently, for a given
density, when $F_\nu$ increases
the thermal equilibrium is reached at lower temperatures
in the regions where $\Lambda_e$ dominates the cooling (high
densities), and the equilibrium curve is
shifted down (see Fig.~1a). In contrast, at low densities
($n \sim 10^{-2}$ cm$^{-3}$), the
dominant cooling mechanism is $\Lambda_{Ly}$,
which decreases when $F_\nu$ increases and, in this case, the
equilibrium is reached at higher temperatures.

Fig.~1a also shows that
two regions of thermal stability, i.e., where the slope
is positive (Field 1965), always exist and two phases
can coexist in
pressure equilibrium if the interstellar pressure $P\equiv p/k$
is between a minimum ($P_{min}$) and a maximum ($P_{max}$) value.
For the case $F_\nu = 0$ we obtain $P_{max} \simeq 1100$ K cm$^{-3}$
and $P_{min} \simeq 490$ K cm$^{-3}$; and if we assume an equilibrium
pressure of $P \simeq 10^3$ K cm$^{-3}$,
then there can be gas with $T \simeq 8800$ K,
$n \simeq 0.1$ cm$^{-3}$ and $n_e/n \simeq 4.7\times10^{-2}$,
and gas with
$T\simeq 100$ K, $n \simeq 10$ cm$^{-3}$ 
and $n_e/n \simeq 1.7\times10^{-3}$ coexisting in equilibrium.
These results
agree roughly with observational estimations of the
warm and cold neutral phases in the local ISM
(Kulkarni \& Heiles 1988).
Furthermore, it can be seen in Fig.~1a that
$P_{max}$, the maximum pressure value over
which only the cold phase can exist, decreases as $F_\nu$ increases.
For $F_{\nu} = 10^2$ cm$^{-2}$ s$^{-1}$ 
and for $F_{\nu} = 10^4$ cm$^{-2}$
s$^{-1}$ we obtain $P_{max} \sim 640$ K cm$^{-3}$ and
$P_{max} \sim 500$ K cm$^{-3}$, respectively;
but observations indicate that the pressure in most of the regions
of the Galactic plane
is $\ga 10^3$ K cm$^{-3}$ (Jenkins et al. 1983). Therefore,
there seems to be an inconsistency between the observed pressure in a
multi-phase medium and high $F_{\nu}$ values. The basic
reason for this behavior (that high $F_{\nu}$ values imply
too low $P_{max}$ values) is that neutrino decay is a poor
heating agent, while other processes can ionize {\it and}
heat the ISM. For instance, when the cosmic ray ionization rate
($\zeta_{cr}$) is changed (keeping $F_{\nu}=0$ fixed), we find
that $P_{max}$ decreases as $\zeta_{cr}$ increases, but
for $\zeta_{cr} \ga 10^{-16}$ s$^{-1}$ the heating by
cosmic rays ($\Gamma_{cr}$) becomes more important that
heating by photoelectrons from grains and
PAHs ($\Gamma_{pe}$), and then
$P_{max}$ begins to increase as $\zeta_{cr}$ is increased. The
minimum $P_{max}$ value is $\sim 900$ K cm$^{-3}$.

However,
more efficient heating mechanisms that those considered here
can rise up the equilibrium curve increasing $P_{max}$. In
order to illustrate this effect, we have plotted in
Fig.~2 $P_{max}$ as a function of $F_{\nu}$ for three
different values of $G_o$ (1, 10 and 20)
and, in consequence, three
different values of $\Gamma_{pe}$ (believed to be an important
heating mechanism in the ISM). We can see that as
$F_{\nu}$ is increased $P_{max}$ decreases until a
minimum value ($\sim 480$ K cm$^{-3}$ for $G_o=1$,
$\sim 590$ K cm$^{-3}$ for $G_o=10$
and
$\sim 660$ K cm$^{-3}$ for $G_o=20$) and after that
remains
constant. This occurs when $n_e/n \rightarrow 1$
in the warm gas and, therefore, additional increases in 
$F_\nu$ do not produce additional changes in this
phase. Fig.~2 shows that if an ISM pressure of
$\sim 10^3$ K cm$^{-3}$ is assumed, a two-phase medium
is possible for $G_o=1$ only if $F_{\nu} \la
10$ cm$^{-2}$ s$^{-1}$, and for $G_o=20$ only if
$F_{\nu} \la 200$ cm$^{-2}$ s$^{-1}$. We conclude
that high fluxes of neutrino decay photons ($\ga 10^3$
cm$^{-2}$ s$^{-1}$) can
be consistent with a two-phase medium only if more
efficient heating sources are acting on the gas.

Sciama (1997a) estimated that $F_{\nu} \sim 3\times10^4$
cm$^{-2}$ s$^{-1}$ is necessary to produce $n_e \sim 0.05$
cm$^{-3}$ at $T \sim 6000$ K. Fig.~3 shows $n_e$ as
function of $F_{\nu}$ for $T = 6000$ K and for the same three
values of $G_o$ given in Fig.~2. The desired
electron density at this temperature can be reached
at high $F_{\nu}$ values
only if $G_o \ga 20$ and, again,
more efficient heating sources seem to be necessary.

An interesting consequence has to be noted: {\it neutrino beams
from neighboring regions may induce the condensation
of cold clouds stimulating the formation of
stars}. The importance of star formation triggered by
previously formed stars has been recognized by many
authors (see the review of Elmegreen 1992). Triggering
mechanisms are usually related to compression of the ISM
by shock waves from close supernovas, because the transition
warm gas $\rightarrow$ cold gas is promoted if the ISM
pressure rises over $P_{max}$. Although this kind of
mechanism can act only over short distances (compared with
the Galaxy size) it can propagate over large scales, and
the idea of self-propagated star formation has been used
to study the formation of spatial patterns in galactic disks
(Mueller \& Arnett 1976; Gerola \& Seiden 1978;
Seiden \& Gerola 1979; Schulman \& Seiden 1990;
Jungwiert \& Palous 1994).
However, the
phase transition warm gas $\rightarrow$ cold clouds
can be also obtained
by decreasing $P_{max}$ under $P$ (assumed constant) if
the local flux of decaying neutrinos increases due, for
example, to an increase in the supernova explosion rate.
This triggering mechanism depends on the
propagation of neutrinos, and therefore can act over large
distances in short time intervals.
On the other hand, star formation
can also inhibit in different ways the warm gas condensation
self-regulating the star formation process
inclusively over large distances (Cox 1983;
Franco \& Shore 1984; Struck-Marcell \& Scalo 1987;
Parravano 1988, 1989).
It has been shown that
star formation inhibition (rather than stimulation) can
also contribute to the formation and maintenance of spatial
patterns in galaxies (Freedman \& Madore 1984; Chappell \& Scalo
1997). Stimulation and inhibition mechanisms of star formation
must be acting simultaneously in the Galaxy,
but it is not clear yet what spatial scales are important for
each one.
The effect of {\it non-local} star formation
stimulation on the formation of spiral patterns in galaxies
should be analysed in future models.

\section{Conclusions}

The thermochemical equilibrium of the ISM, including decay of
neutrinos into an ionizing photon flux $F_{\nu}$, was calculated.
The range $0 \leq F_{\nu} \leq 10^5$ cm$^{-2}$ s$^{-1}$ always
shows two regions of stability (a warm and a cold phase)
that can coexist in equilibrium if the ISM pressure is below
a threshold value ($P_{max}$). High $F_{\nu}$ values
($\sim 3\times10^4$ cm$^{-2}$ s$^{-1}$)
estimated by Sciama (1997a) to produce $n_e \simeq 0.05$
cm$^{-3}$ at $T\simeq6000$ K, only can be consistent with
observed ISM pressures if more efficient processes are heating
the gas.
It also was showed that a neutrino flux increase (due, for example,
to an increase in the supernova explosion rate)
may stimulate the condensation of cold gas (and probably the
star formation) by decreasing $P_{max}$ under the ISM pressure
value.

\begin{acknowledgements}

This work has been partially supported by CONDES
of Universidad del Zulia.
The authors are very grateful to an anonymous referee
for several helpful suggestions, and
to Cesar Mendoza for
his assistance in the manuscript preparation.

\end{acknowledgements}

\clearpage

\centerline{{\bf Figure captions}}
\begin{itemize}
\item[{\bf Fig. 1a.}] Thermal pressure as a function of the total
hydrogen density in equilibrium for $G_o=1$ and for
$F_\nu = 0$ (solid line),
$F_\nu = 10^2$ (dashed line) and
$F_\nu = 10^4$ cm$^{-2}$ s$^{-1}$ (dot-dashed line).
\item[{\bf Fig. 1b.}] The electron fraction as a function of the total
hydrogen density in equilibrium for $G_o=1$ and for
$F_\nu = 0$ (solid line),
$F_\nu = 10^2$ (dashed line) and
$F_\nu = 10^4$ cm$^{-2}$ s$^{-1}$ (dot-dashed line).
\item[{\bf Fig. 2.}] The maximum pressure $P_{max}$ for the coexistence
of warm and cold gas as a function of
$F_\nu$ for
$G_o=1$ (solid line),
$G_o=10$ (dashed line)
and $G_o=20$ (dot-dashed line).
\item[{\bf Fig. 3.}] The electron density as a function $F_\nu$ for
$T=6000$ K and for
$G_o=1$ (solid line),
$G_o=10$ (dashed line)
and $G_o=20$ (dot-dashed line).
\end{itemize}

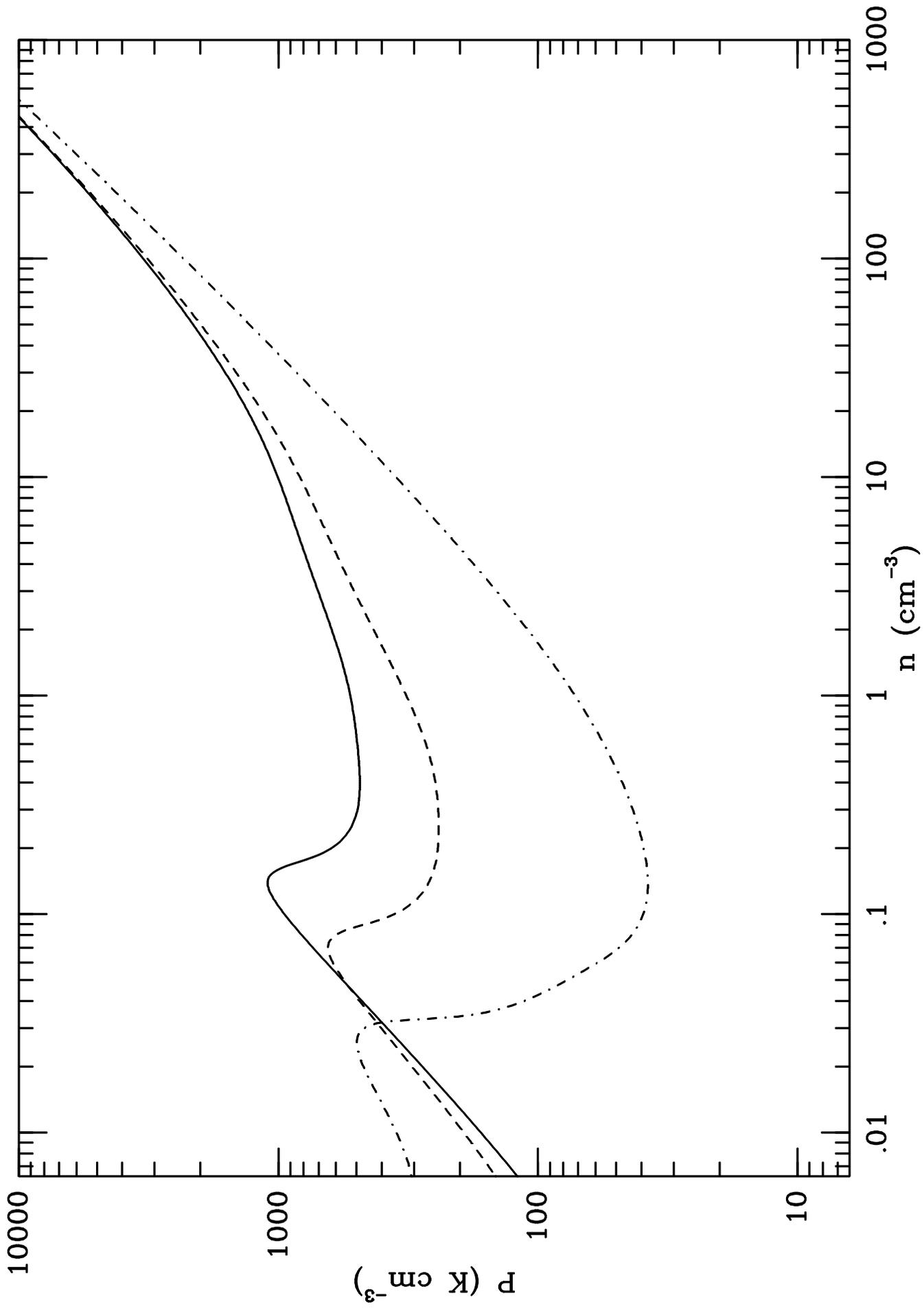
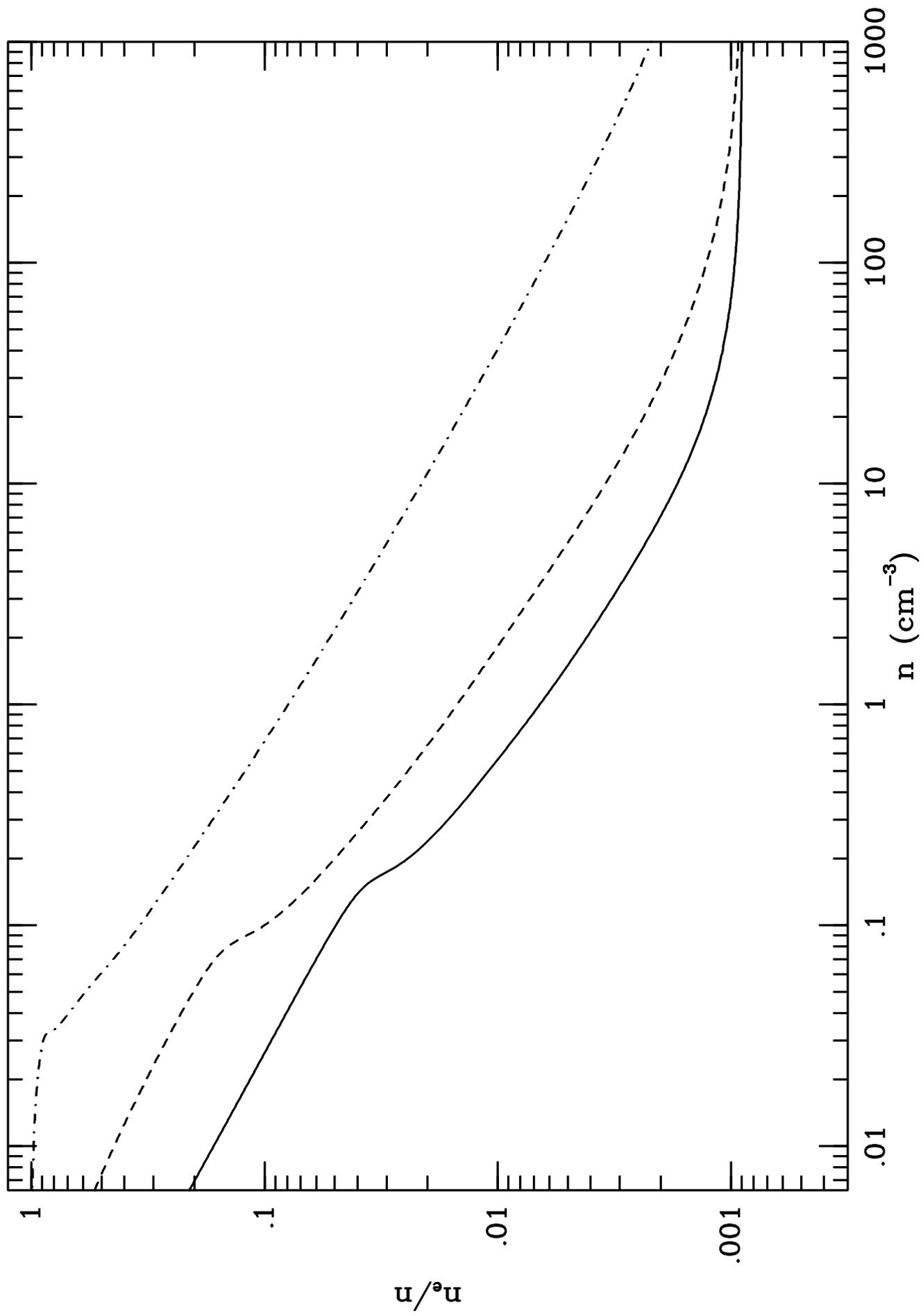
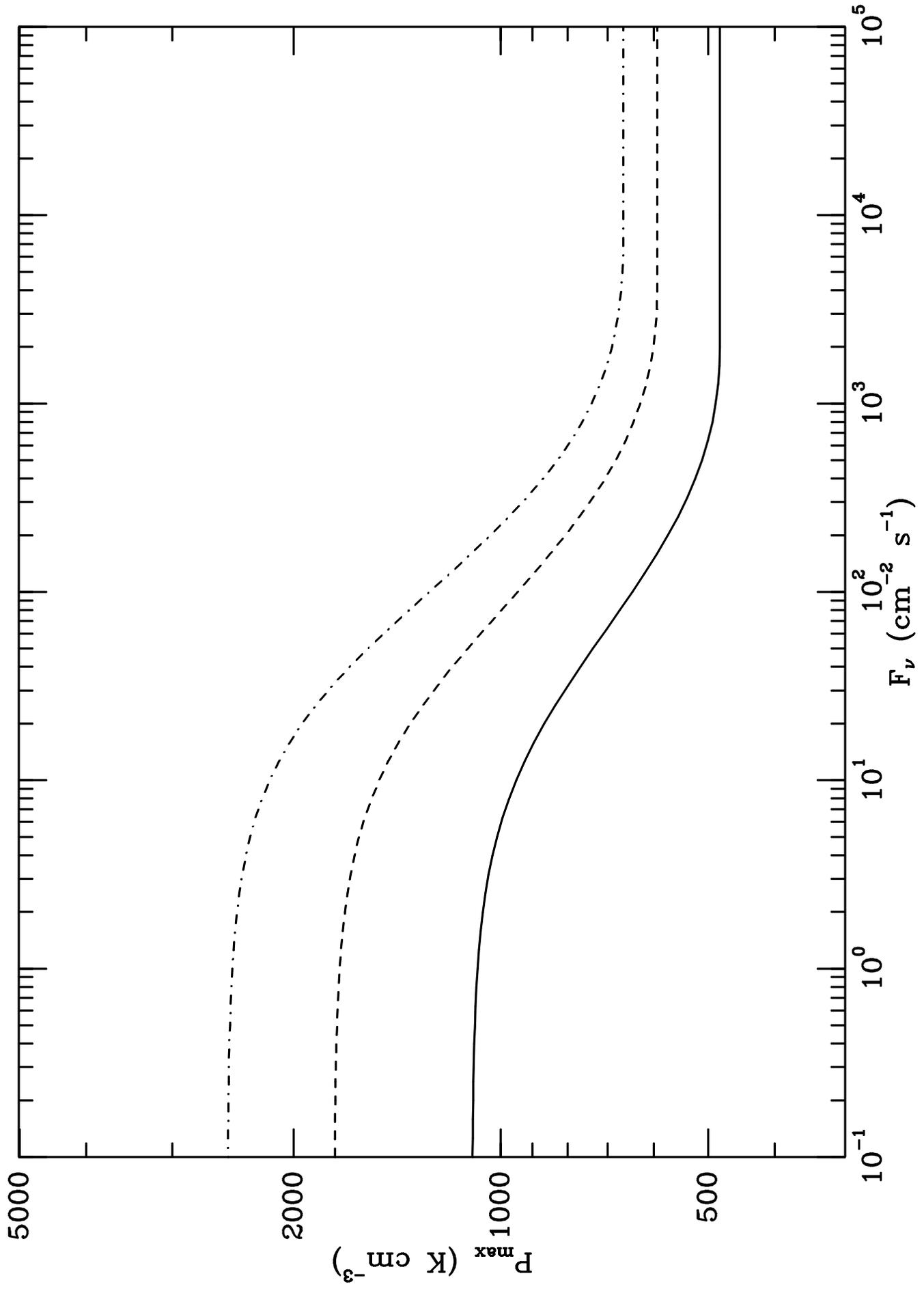
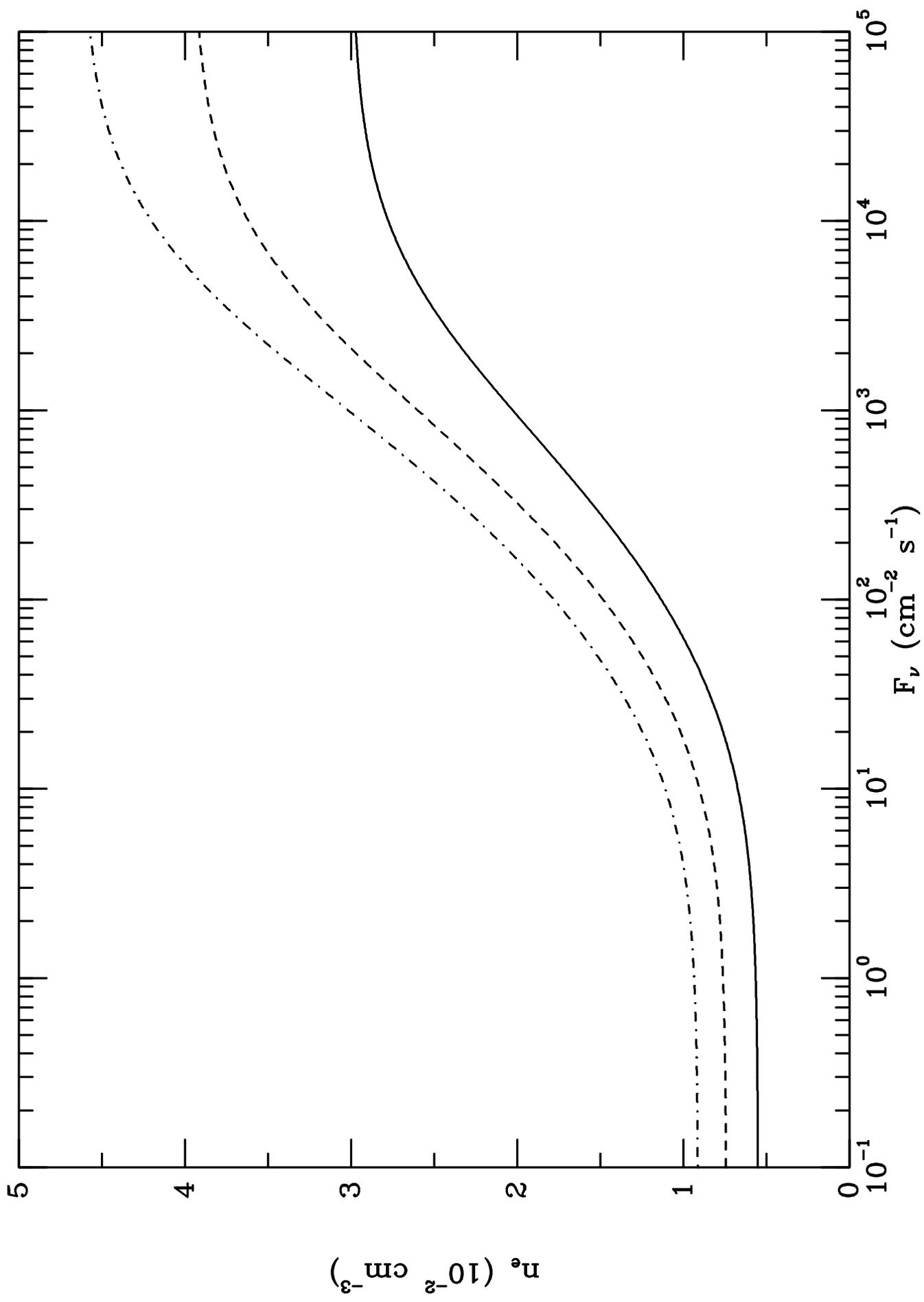
\end{document}